# Chasing the 'killer' phonon mode for the rational design of low disorder, high mobility molecular semiconductors

Guillaume Schweicher[†][*][1], Gabriele D'Avino[†][2], Michael T. Ruggiero[†][3,4], David J. Harkin[1], Katharina Broch[1], Deepak Venkateshvaran[1], Guoming Liu[1], Audrey Richard[5], Christian Ruzié[5], Jeff Armstrong[6], Alan R. Kennedy[7], Kenneth Shankland[8], Kazuo Takimiya[9], Yves H. Geerts[5], J. Axel Zeitler[3], Simone Fratini[*][2], Henning Sirringhaus[*][1]

[1] Optoelectronics Group, Cavendish Laboratory, University of Cambridge, JJ Thomson Avenue, Cambridge CB3 0HE, United-Kingdom

[2] Institut Néel-CNRS and Université Grenoble Alpes, Boîte Postale 166, F-38042 Grenoble Cedex 9, France

[3] Department of Chemical Engineering and Biotechnology, University of Cambridge, Philippa Fawcett Drive, Cambridge, CB3 0AS, United-Kingdom

[4] Department of Chemistry, University of Vermont, 82 University Place, Burlington, VT 05405, United States of America

[5] Laboratoire de Chimie des Polymères, Faculté des Sciences, Université Libre de Bruxelles (ULB), Boulevard du Triomphe CP206/01, 1050 Brussels, Belgium

[6] ISIS Facility, Rutherford Appleton Laboratory, Harwell Oxford, Didcot, Oxfordshire OX11 0QX, United-Kingdom

[7] Department of Pure and Applied Chemistry, University of Strathclyde, 295 Cathedral Street, Glasgow G1 1XL, Scotland

[8] School of Pharmacy, University of Reading, Whiteknights, Reading RG6 6AD, United-Kingdom

[9] Emergent Molecular Function Research Group, RIKEN Center for Emergent Matter Science (CEMS), Wako, Saitama, Japan

[†]These authors contributed equally to this work.
*e-mail: gs524@cam.ac.uk; simone.fratini@neel.cnrs.fr; hs220@cam.ac.uk

**Abstract**

Molecular vibrations play a critical role in the charge transport properties of weakly van der Waals bonded organic semiconductors. To understand which specific phonon modes contribute most strongly to the electron – phonon coupling and ensuing thermal energetic disorder in some of the most widely studied high mobility molecular semiconductors, we have combined state-of-the-art quantum mechanical simulations of the vibrational modes and the ensuing electron phonon coupling constants with experimental measurements of the low-frequency vibrations using inelastic neutron scattering and terahertz time-domain spectroscopy. In this way we have been able to identify the long-axis sliding motion as a 'killer' phonon mode, which in some molecules contributes more than 80% to the total thermal disorder. Based on this insight, we propose a way to rationalize mobility trends between different materials and derive important molecular design guidelines for new high mobility molecular semiconductors.

**Main Text:**

The last decade has witnessed drastic improvements of the electronic properties, environmental and operational stability and processibility of organic semiconductors (OSCs) [1,2]. Designing new materials with high carrier mobilities, µ, remains one of the main research objectives to enable faster operation and lower power consumption of circuits and addressing of advanced liquid crystal and organic light-emitting diode (OLED) displays[1,3]. Yet despite exploring a wide range of material systems, charge carrier mobilities in excess of 10 cm$^2$ V$^{-1}$ s$^{-1}$ have only been achieved in very few molecular semiconductors and highly aligned polymers[4–6]. At present, despite significant general advances in the comprehension of transport physics, a truly microscopic understanding is still missing and the search for high mobility materials often remains heuristic.

The complex charge transport mechanism taking place in weakly, van-der-Waals bonded molecular crystals has been a topic of intense debate[3,7]. Clear descriptions of transport exist in the limits of very low (µ << 1 cm$^2$/Vs) and very high charge carrier mobilities (µ >> 10 cm$^2$/Vs), where transport can be described in terms of incoherent hopping[8] of localized charges between sites and Boltzmann band transport of delocalized Bloch electrons, respectively[9]. However, at present for most state-of-the-art molecular semiconductors mobilities are in an intermediate range, where hopping transport is no longer applicable, but mobilities at room temperature remain too small to satisfy one of the main criteria for Boltzmann transport; the mean free path for carrier scattering with lattice phonons has to be significantly larger than the lattice constant. In recent years, detailed theories of transport for this intermediate regime have been developed that take into account the key characteristics of molecular solids: Due to the soft, van der Waals intermolecular bonding the phonon modes in molecular crystals tend to exhibit large vibration amplitudes on the order of 0.1Å[10,11]. These thermal lattice fluctuations cause temporal variations in the transfer integrals and site energies across the molecular lattice. This so-called dynamic disorder limits the ability of charge carriers to form fully delocalized Bloch electron states and imposes a transient localization on the charges leading to a unique transport regime, in which the carriers exhibit both localized and extended characters[12–15]: On a timescale shorter than the characteristic timescale for intermolecular vibrations the charges are localized by the thermal disorder in the transfer integrals, but on longer timescales the charge carriers undergo diffusive motion driven by the waves of molecular lattice fluctuations. This transient localization scenario is able to reconcile many of the experimental features of transport, including the concomitant observation of delocalized transport signatures, such as an ideal Hall effect and an increase of the coherence factor with decreasing temperature and increasing pressure[16–19], and localized signatures, such as radical-cation like charge-induced optical absorptions[20] or strong electron-phonon coupling effects in their band dispersion[20,21]. It appears also able to reproduce the observed temperature dependence of the mobility[14], although experimentally it is difficult to establish in many systems that temperature dependent mobility measurements are not affected by extrinsic factors, such as contact resistance, presence of traps or mechanical stress[22].

Within the transient localization scenario[15], the charge transport is governed by few key parameters: To achieve the highest carrier mobilities one should minimize the amount of thermal molecular disorder, and at the same time reduce the sensitivity of carrier motion to such disorder. The second condition can be realized with a proper engineering of the inter-molecular overlaps: as recently shown, isotropic band structures with approximately equal transfer integrals in the different crystal directions show the highest resilience to dynamic disorder, as they minimise localization phenomena[15]. Whether this requirement is satisfied in a particular system or not, can be established once the crystal structure is known. However, to achieve the condition of low thermal disorder no effective strategies are currently known apart from the generic requirement for stiff structures in order to avoid large-amplitude vibrational modes[11,19]. Little is known to which extent there is scope to influence the thermal disorder by molecular design. The difficulty arises because of the complex unit cell of molecular crystals, which typically contain on the order of 100 atoms. This means that there are hundreds of phonon modes which can in principle couple to the charge motion and contribute to the thermal disorder. Do all these modes make important contributions or are there specific 'killer' modes that are responsible for generating most of the thermal disorder? In the former case the complexity would be bewildering and there would be little hope for any systematic molecular design to suppress thermal disorder. However, in the latter case we will need to understand the nature of these 'killer' modes in order to derive molecular design guidelines. A detailed, mode-resolved understanding of the vibrational dynamics and electron-phonon coupling, in particular of low-energy intermolecular vibrations, that are more strongly excited than intramolecular modes, is required for the rational design of the next generation of OSCs.

Low frequency spectroscopy techniques have long been used to investigate phonon modes in molecular solids, for example, in the context of pharmaceutical drug development[23–27] or for identifying polymorphism in OSCs[28–30]. Quantum mechanical simulations provide powerful tools to interpret the modes observed in experiments; considerable emphasis has been spent over the last decade to achieve the necessary computational accuracy required to fully capture the very weak forces that are responsible for the low-frequency motions in van der Waals bonded molecular crystals and are more challenging to model than in more strongly bonded systems. [24,25]. Here we use this state-of-the-art methodology to compare the vibrational dynamics of several important classes of high mobility molecular semiconductors. To experimentally determine the molecular frequencies we use terahertz time-domain spectroscopy (THz-TDS) and inelastic neutron scattering (INS). We model the spectra by fully-periodic quantum mechanical calculations and use the obtained set of normal phonon modes and their frequencies to determine the vibrational amplitudes and the corresponding mode-resolved electron-phonon coupling constants (see SI section S4 and S5). This methodology enables us to assess the contributions that individual modes make to the total thermal disorder $\sigma$, which can be understood as a sum over the electron phonon couplings of the individual modes weighted with their amplitudes. Similar approaches have very recently been reported by Tu[31] and Harrelson[32]. Here our focus is on applying this mode-resolved analysis to assess whether there are particular phonon modes that dominate the thermal disorder and understand the impact of any such 'killer' modes on carrier mobilities calculated within transient localization theory.

We focus our work on the classical high mobility OSCs, pentacene and rubrene, as well as a group of thienoacenes, which have emerged as a very promising class of high mobility molecules. Brickwall materials (TIPS-Pentacene, diF-TES-ADT,...) have not been investigated here because of their non-ideal dimensionality of charge transport, lower experimental mobilities and larger molecular motions[3]. Figure 1 highlights the full set of studied materials and their respective molecular packing. BTBT and DNTT, as well as their alkylated derivatives C8-BTBT-C8 and C8-DNTT-C8, all exhibit a herringbone packing similar to that of pentacene, while rubrene, the highest mobility ($\mu$) molecule recorded so far ($\mu \cong 20$ cm$^2$ V$^{-1}$ s$^{-1}$), exhibits a 1D slipped-cofacial packing[6]. Colored arrows indicate the inequivalent pairs of molecules, which give rise to different transfer integral values within the unit cell. The figure also shows an illustrative example of the characteristics of polycrystalline, bottom-gate, top-contact organic field-effect transistors (OFETs) of DNTT and C8-DNTT-C8. The devices exhibit near ideal characteristics and allow a reliable extraction of mobility values. We consistently observe higher mobilities in the alkylated species ($\mu = 3.4$ cm$^2$ V$^{-1}$ s$^{-1}$) compared to the non-alkylated one ($\mu = 1.4$ cm$^2$ V$^{-1}$ s$^{-1}$) (see SI Section S2 for detailed information). A similar trend has also been reported in the literature for more optimized single-crystal devices with even higher mobilities[33,34]. This is an example of significant mobility differences between related molecular systems, which are commonly reported in the literature, but are currently not well understood. Our main aim in this paper is to investigate to which extent differences such as this can be explained fully in terms of the isotropy of the band structure, as previously suggested[15], or whether they also reflect important differences in the mode-resolved electron phonon couplings and resulting thermal disorder.

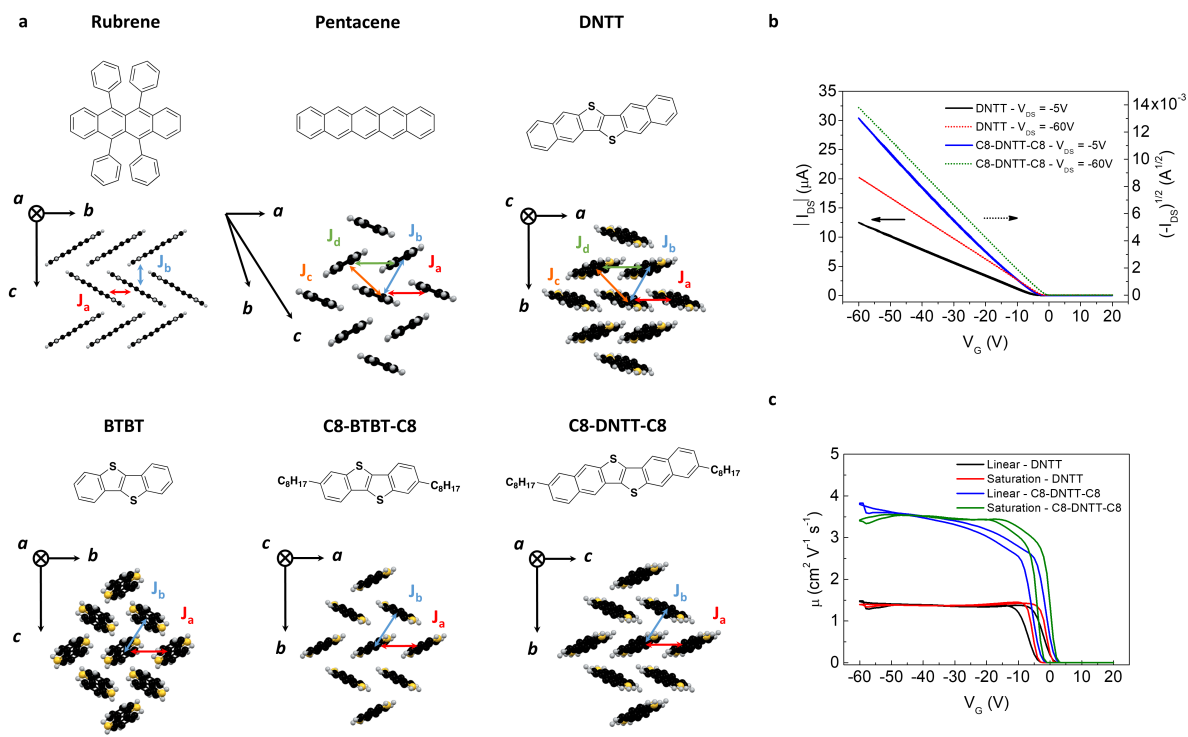

**Figure 1. Molecular packing of investigated materials and thin-film OFET characteristics of DNTT and C8-DNTT-C8.** (a) Molecular structure and associated top

view of the respective packing structure of all the molecules investigated (side chains and rings are omitted for clarity); (b) Transfer characteristics in the linear and saturation regime for DNTT and C8-DNTT-C8 OFETs (L = 350 μm, W = 1 mm, for the linear (saturation) characteristics the current (square root of the current) is plotted on the left (right) axis; **(c)** Corresponding gate-voltage dependence of the extracted linear and saturation mobility.

The THz-TDS measurements were performed at 80 K, while INS measurements were performed at around 10K at the ISIS TOSCA beamline (Harwell, UK). The two measurements complement each other: INS is able to cover a broad range of frequencies, routinely from 150 cm$^{-1}$ to the mid-IR (1500 cm$^{-1}$), and has the advantage that the neutron scattering cross section does not involve any mode-specific selection rules. For the interpretation of THz-TDS measurements optical selection rules need to be considered, but they extend the accessible range to very low frequencies (20 cm$^{-1}$ in our setup). The spectra (Figure 2) reveal the existence of many low-frequency modes in all the investigated solids; the alkylated C8-BTBT-C8 and C8-DNTT-C8 molecules exhibit an even larger number of modes than the unsubstituted ones due to the increased number of atoms in the crystallographic unit cell. Unlike mid-infrared measurements, where characteristic intramolecular modes can often be attributed to specific functional groups and differences in molecular design or crystal packing lead to slight frequency shifts, low-frequency measurements instead probe the complex potential energy landscape and weak intermolecular interactions of the entire molecular crystal. As such, they require a careful theoretical analysis by solid-state density functional theory (DFT). We have hence calculated the harmonic lattice dynamics of the selected molecular crystals with accurate, all-electron DFT simulations. Following a very recent benchmark study[35], we employed the D3-corrected[36,37] Perdew-Burke-Ernzerhof[38] (PBE) functional, which has been shown to outperform other existing approaches in the description of the subtle interplay of covalent and non-covalent interactions in complex molecular crystals (for details see SI section S4). One limitation of our technique is that in the THz-TDS experiments and in the simulations, it was possible to measure and compute phonons only at the Brillouin zone centre; our INS measurements at TOSCA sample over the entire Brillouin zone, although at energies > 200 cm$^{-1}$ the phonon dispersion is expected to be almost negligible[39]. In principle INS is capable of resolving phonon dispersion, but this would have required large single crystals with a weight on the order of 1g, which unfortunately cannot be grown for these materials.

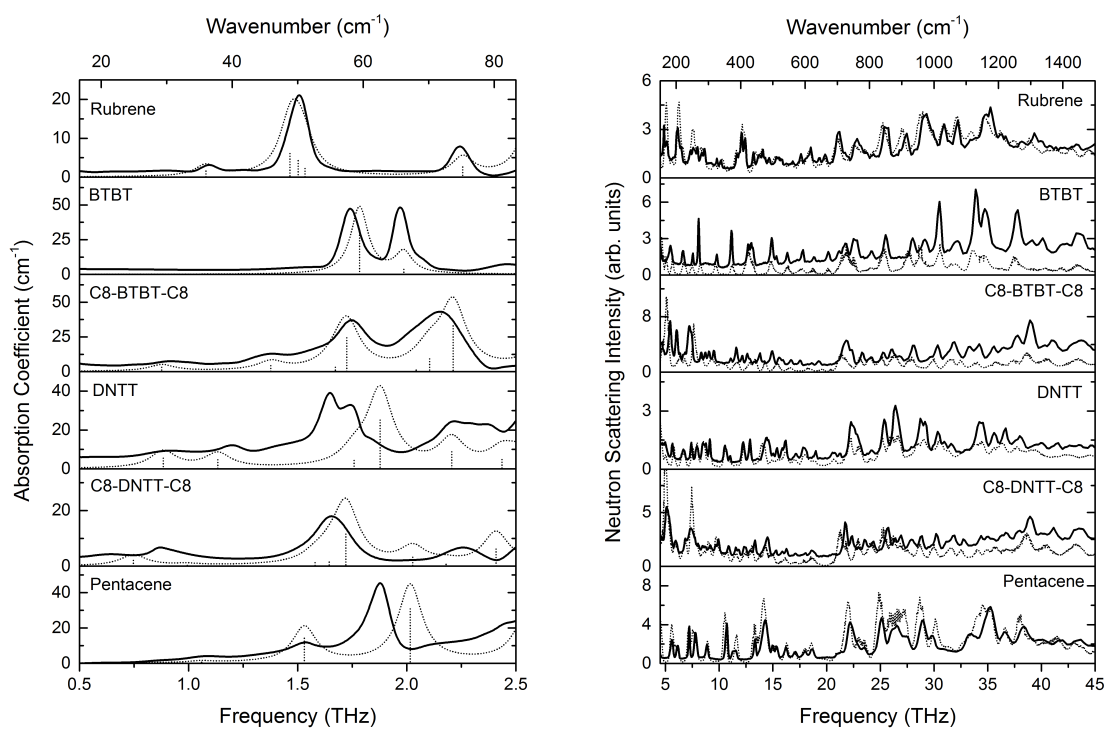

**Figure 2. Experimental characterisation of low-frequency intermolecular vibrations.** (left) THz-TDS spectra and (right) INS spectra. The experimental data are shown as solid black lines and respective DFT-simulated spectra are shown as grey dashed-lines. A Lorentzian broadening using the empirical full-width at half maximum of the modes has been applied to the simulated THz-TDS spectra.

The results of our normal mode calculations (dashed lines in Figure 2) adequately reproduce the experimental THz-TDS and INS spectra, both in terms of vibrational frequencies (with a maximum discrepancy to experiments of 7 cm$^{-1}$) and relative peak intensities. The comparison with the experimental data provides an important validation step for the calculated frequencies and atomic displacements of the normal modes that will be used in the following for the assessment of electron-phonon couplings. Our harmonic lattice dynamics simulations for the unsubstituted molecules do not predict any optically-active vibrational mode at frequencies below the window accessible in THz spectroscopy. However, a number of ultra-soft normal modes (frequencies well below 0.5 THz) are found in alkylated systems. A close inspection of the potential associated with these modes reveals a pronounced anharmonic character, which plausibly originates from the mixing of nearly rigid molecular displacements with the, notably anharmonic, torsions of the side chains. Accurate estimates of the vibrational frequencies for these modes were hence obtained through the exact numerical solution of the Schrödinger equation associated with the anharmonic *ab initio* potential. Anharmonic modes were found to be substantially harder than what predicted within the harmonic approximation (for further details on anharmonic calculations see SI Section S4). All modes above 1 THz were well reproduced by the harmonic approximation and were analysed accordingly.

Having calculated and validated the full spectrum of vibrational modes, we can now establish comprehensive mode-resolved electron-phonon coupling maps for each one of the different materials under study. This represents a necessary step in any theory of phonon-limited electronic transport, which can be either formulated in reciprocal space as customary in Bloch-Boltzmann transport[9,40] or in a real-space picture as we do here. While the two approaches are perfectly equivalent, we opt for the latter as it provides a more insightful framework for establishing structure-property relationships in molecular semiconductors[41,42]. These maps consist of the off-diagonal electron-phonon couplings $\beta_{xm} = \partial J_x / \partial q_m$, reflecting the modulations induced by phonon mode $q_m$ of the transfer integral $J_x$ between pairs of molecular sites labelled by the index $x$ (see Figure 1), as well as the intra-molecular (diagonal) couplings $\alpha_{im} = \partial \varepsilon_i / \partial q_m$, corresponding to modulations of the site energy $\varepsilon_i$ of the $i$-th molecular site. We shall focus primarily on the off-diagonal couplings to low-frequency modes, as these are the most detrimental for charge transport (the calculated maps for diagonal couplings are presented in SI section S5).

We first focus on the comparison of DNTT with its alkylated version C8-DNTT-C8. Figures 3(a,e) compare the calculated amplitudes $\sigma_q$ of all the zone centre vibrational modes in the 0-400 cm$^{-1}$ frequency range. In addition to the aforementioned increase in the number of normal modes upon alkylation, the calculations clearly illustrate the decrease in vibrational amplitude with frequency expected for classical harmonic oscillators at thermodynamic equilibrium ($\sigma_q^2 \propto T/\omega^2$). Figures 3(b,f) compare the calculated values of the off-diagonal couplings $\beta_{cm}$ (for $x = c$ these couplings are the largest, equivalent maps for the other molecular pairs are reported in SI section S5). In both materials there are a few specific modes across the spectrum that exhibit particularly large couplings and could in principle have a strong impact on charge transport, while most modes exhibit only weak coupling and are likely to be less relevant.

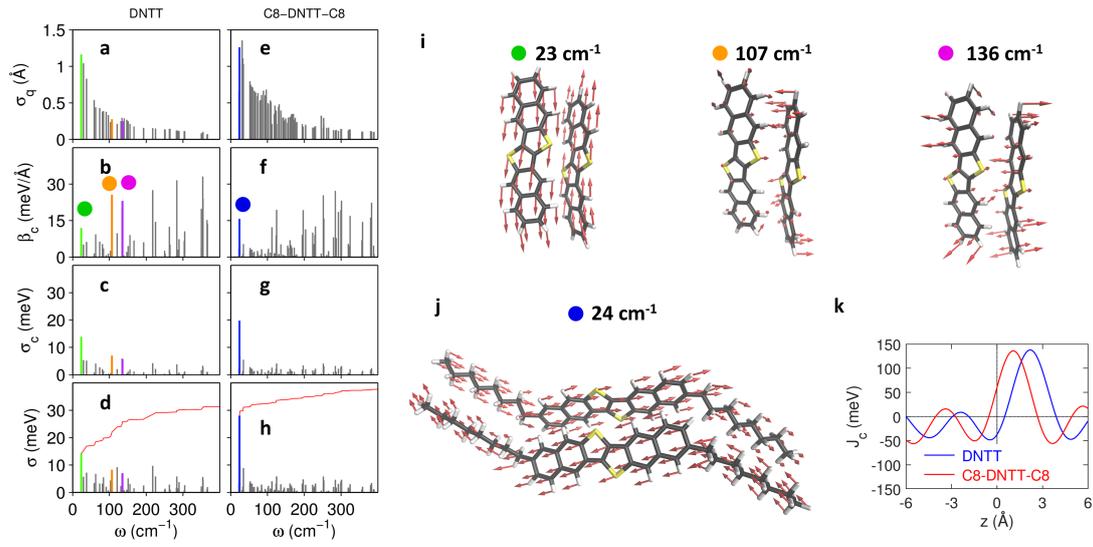

**Figure 3. Electron-phonon coupling maps and identification of most detrimental modes – Comparison between DNTT and C8-DNTT-C8.** **(a,e)** Amplitudes $\sigma_q$ of the zone centre vibrational modes as a function of their frequency for DNTT (a) and C8-DNTT-C8 (e); **(b,f)** Corresponding, mode-resolved non-local electron-phonon coupling (for a specific dimer ($x = c$); **(c,g)** Corresponding mode-resolved fluctuation of the transfer integrals $\sigma_c$. Specific modes are identified that make the strongest contributions to the transfer integral. Examples include modes at 23 cm$^{-1}$, 107 cm$^{-1}$, 136 cm$^{-1}$ for DNTT and 24 cm$^{-1}$ for C8-DNTT-C8; **(d,h)** Mode-resolved fluctuations of the transfer integrals averaged over all inequivalent pairs of molecules in the unit cell together with a cumulant plot (red line). All fluctuations are evaluated at room temperature; **(i-j)** Relative displacements of neighbouring molecules associated with specific modes that contribute strongly to the energetic disorder (for video representations of these modes see SI). **(k)** Transfer integral $J_c$ of a single molecular pair as a function of the sliding coordinate for DNTT and C8-DNTT-C8.

From the quantities presented above we can now evaluate how each of the vibrational modes contributes to the energetic disorder, and therefore track which modes are more detrimental to charge transport. The fluctuations induced by each mode on a given transfer integral (for example, $J_c$) are actually determined by the product $\sigma_c = \sigma_q \beta_c$ of the mode amplitude $\sigma_q$ and its vibrational coupling $\beta_c$, plotted in Figure 3(c,g). To assess the total vibrational disorder in a given material, one then has to sum up the contributions of all modes to the different intermolecular transfer integrals, and consider the total fluctuation $\sigma = \sqrt{\sigma_a^2 + \sigma_b^2 + \sigma_c^2}$ (Figure 3(d,h), see SI section S5 for details). From the data we can clearly identify a few dominant, low-frequency modes which contribute most strongly to the total disorder, i.e. those corresponding to the most prominent bars and the most prominent steps in the cumulant plot (red line). These are the modes which combine large amplitude and large coupling. In both molecules the most detrimental vibration is a low frequency phonon at 23 cm$^{-1}$ (DNTT) and 24 cm$^{-1}$ (C8-DNTT-C8), which involves mainly long-axis relative displacements of neighboring molecules in opposite directions (see Figure 3(i) and (j), respectively). Strikingly, this single "sliding mode" is responsible *per se* for a conspicuous fraction of the total transfer integral fluctuations, ranging from 45% in DNTT to as much as 75% in C8-DNTT-C8; this justifies referring to it as the 'killer' mode. In the case of DNTT, the remaining energetic disorder can mostly be attributed to few other "in-plane modes" at higher frequency, typically below 150 cm$^{-1}$, mostly implying displacements of the core atoms within the herringbone layers (see Figure 3(i)). Modes above 150 cm$^{-1}$ do not contribute significantly owing to their negligible vibrational amplitude.

Based on our *ab initio* modeling, the effects of alkylation can therefore be understood as follows: (i) new modes emerge in the relevant frequency window; these however mostly involve the alkyl side chains and not the functional cores, and are thus harmless to charge transport; (ii) in-plane modes are shifted to higher frequencies, hence suppressing their impact on disorder and resulting in energy fluctuations becoming dominated by a single sliding mode; (iii) a detailed analysis, presented in the SI, shows that the presence of side chains reduces the amplitude of out-of-plane sliding motion, in accordance with what was reported by Illig et al.[10]. Surprisingly, while all these effects taken together should lead to a reduction of disorder, we find that (iv) the total disorder *increases* upon alkylation (red lines in Figure 3(d,h)). This counterintuitive behavior can be understood by studying the transfer integral of individual molecular pairs as a function of the relative intermolecular shift along the molecular long axis (sliding coordinate *z*), as illustrated in Figure 3(k) and SI Figure S18. Our finding that the long-axis sliding motion is the 'killer' mode indeed justifies a simple analysis with the relevant quantity determining the thermal disorder being the derivatives $dJ_x/dz$ at the equilibrium position ($z = 0$). While the curves for DNTT and C8-DNTT-C8 in Figure 3(k) are very similar in shape, the equilibrium position in DNTT is very close to a "sweet spot" where $J_c(z)$ attains an extremum, implying small values of the derivative (this situation is analogous to that of rubrene[43], which is known to feature a particularly low value of disorder). Alkylation moves the system out of this sweet spot, increasing the coupling by a factor of ~2, ultimately determining the larger energetic disorder with respect to DNTT.

The protocol devised above can now be extended to the whole set of compounds studied in this work. Figure 4 shows the cumulative plots of the transfer integral fluctuations $\sigma$ in units of $J = \sqrt{J_a^2 + J_b^2 + J_c^2}$, the latter setting the overall scale of the electronic bandwidth[15]. The different systems can be classified into three distinct classes: (1) Alkylated systems (b,d) feature a single vibration dominating the energetic disorder --- the sliding mode --- responsible for the large initial step in the cumulative plots of σ. (2) Non substituted systems presenting a pure herringbone packing, such as DNTT, BTBT and pentacene (a,c,f) present a much reduced disorder from the sliding mode, as they all lie close to a sweet spot; the remaining disorder is distributed among a small number of in-plane vibrations (3) Rubrene

(e) singles out because of its peculiar molecular shape, where phenyl rings orthogonal to the backbone act to inhibit any possible sliding motion. This is reflected in the absence of a low-frequency step-like feature in the cumulative disorder plots. Note that the sliding mode for the molecular pairs *a* in Figure 1(a), which is a zone-boundary phonon, is not included in our analysis based on Gamma-point vibrations; this leads to an underestimation of $\sigma_a$. Our analysis here shows that a refined treatment that takes into account zone-centre phonons plus the specific zone-boundary phonons that lead to an asymmetric sliding motion of the *a* molecular pairs would recover most of the energetic disorder.

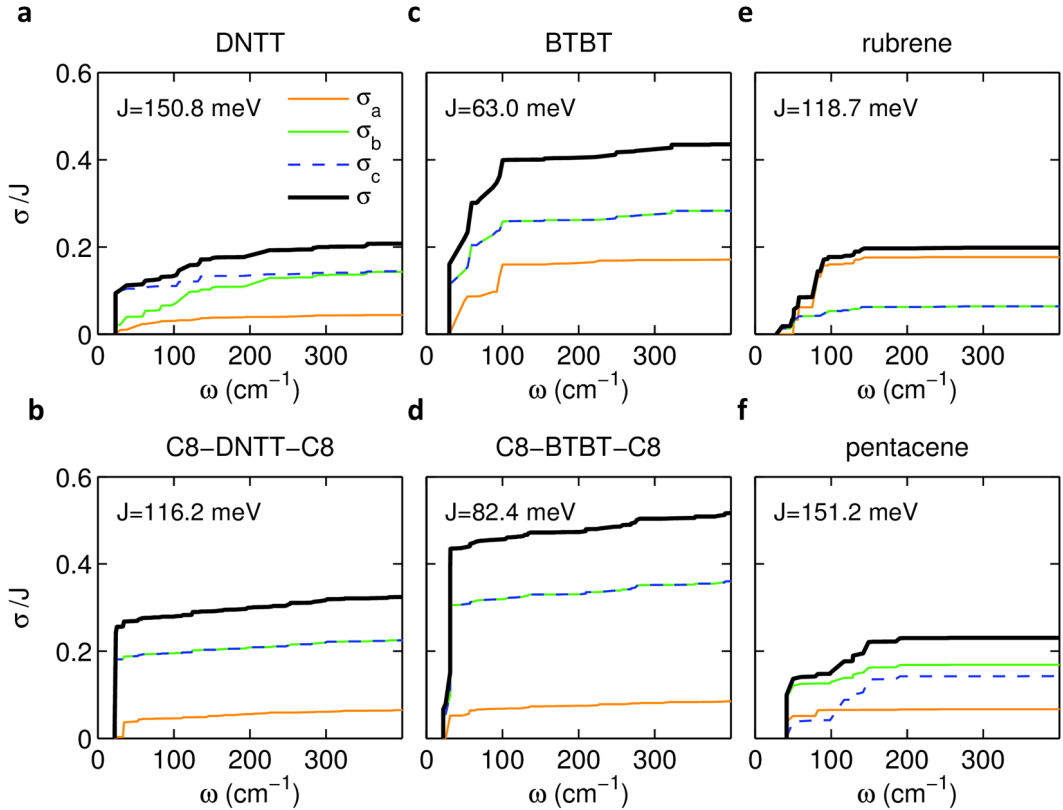

**Figure 4. Comparison of total electron-phonon coupling maps between the investigated materials.** The different lines show the cumulative room-temperature standard deviation of the transfer integral fluctuations normalized by the transfer integral for each of the inequivalent pairs of molecules in the unit cell ($x = a$ (red), $b$ (green), $c$ (blue)) as well as the averaged normalized transfer integral fluctuation $\sigma/J$ which determines the overall degree of energetic disorder. The blue and green lines are identical for BTBT, rubrene, C8-DNTT-C8 and C8-BTBT-C8 as these molecules only have two inequivalent pairs of molecules in their unit cells.

The comprehensive picture of mode-resolved build-up of energetic disorder established above provides us with the missing key ingredient needed to explain the experimental mobility trends in different classes of molecular crystals. To demonstrate this we now incorporate our results into the framework of transient localization theory, where the mobility was recently shown to depend on two dominant factors[15]: (a) the anisotropy and sign of transfer integrals $J_x$ (see SI Table S5), that determine how a given OSC responds to disorder; the mobility in the most conducting direction can increase by up to a factor of 6 when going from a purely one-dimensional to a fully isotropic structure; (b) the magnitude of the energetic disorder, the effect of which has not been considered in detail yet. Importantly, we show below that the dimensionless parameter $\sigma/J$ as used in Figure 4 above provides an effective way to quantify the energetic disorder of a given material and enables us to rationalize differences in mobility between materials. $\sigma/J$ fully determines the effect of disorder on the transport properties, with the mobility expected to scale as $\mu \propto (\sigma/J)^{-\gamma}$, with $\gamma \approx 2$. According to this simple rule, the variations in disorder reported in Figure 4, $\sigma/J = 0.2 - 0.5$, entails variations of a factor of 6 in the mobility, totally comparable with the effects of anisotropy. It is clear then that anisotropy and disorder compete at the same level, and both aspects need to be considered in order to capture the experimental trends.

We report in Figure 5 the measured mobilities for the different molecules in the set, and compare them with a full calculation of the mobility using transient localization theory[15], based on the parameters in SI tables S5 and S7. For the experimental values we have selected the most reliable[44,45] reported mobilities based on robust four-point-probe measurements and further validated by Hall effect measurements (see SI Table S3), which was possible for all compounds except BTBT, due to its high contact resistance. For the theoretical mobilities we show values calculated along the highest mobility direction, taking into account both off-diagonal and diagonal energetic disorder (methods section 7). The data show a good correlation with the experiments, especially considering the large/known/documented sample-to-sample variability characterising organic semiconductors. Quite expectedly the actual experimental values are lower than predicted, because the calculations disregard extrinsic effects that limit the experimentally achievable mobility values, such as contact resistance limitations, chemical impurities and structural defects. C8-BTBT-C8 singles out as having a mobility that is higher than the main trend, actually very close to the theoretical value; this may be because this molecule is known to have self-assembling, liquid crystalline properties, which facilitate growth of thin films with a very low number of extrinsic defects.[46–48] Also, as mentioned above, our simulations do not take into account the dispersion of phonon frequencies away from the Brillouin zone centre, in particular the effect of the zone boundary phonon that involves a sliding motion of the $a$ molecular pair, this is expected to increase the amount of disorder, hence lowering further the theoretically predicted values as recently reported.[31,32,49]

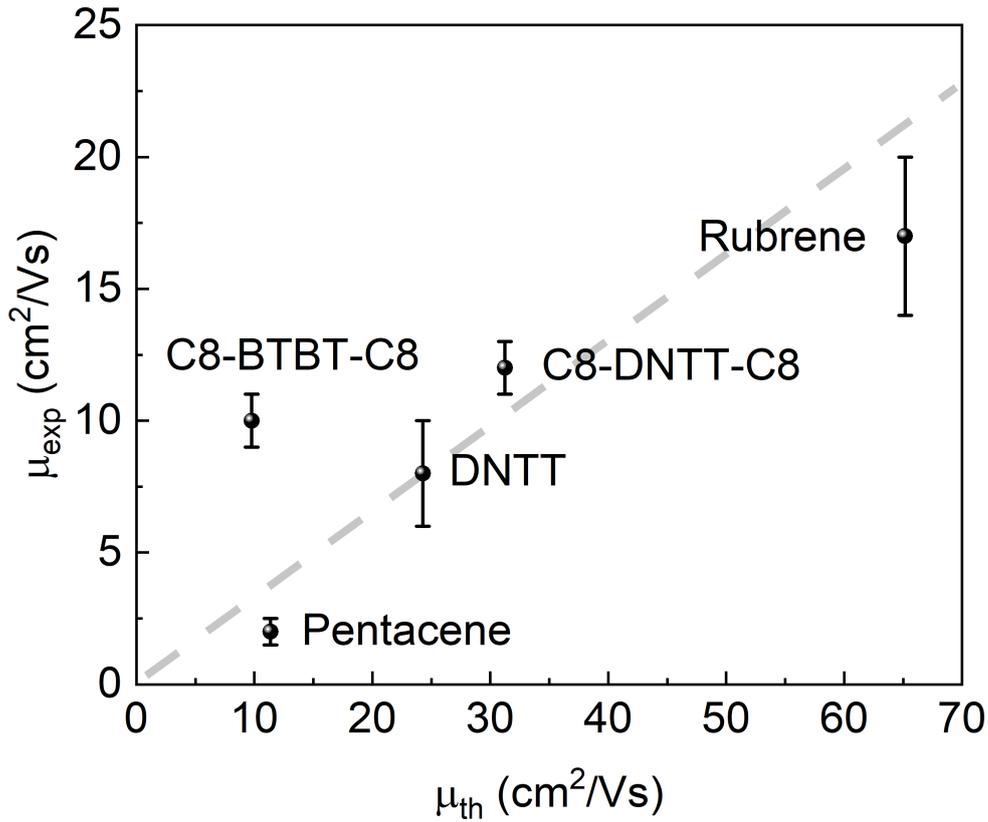

**Figure 5. Comparison between experimentally measured and theoretically predicted mobility values within the transient localization framework.** The theoretical mobility values in the highest mobility direction are compared against the experimental values; We show theoretical predictions including both off-diagonal and diagonal electron-phonon coupling. Energetic disorder is calculated accounting for all low-frequency modes up to 300 cm$^{-1}$. The grey shaded area is a guide to the eye that highlights the correlation between experimental and theoretical values.

We can now understand how the effects of the structure on the inter-molecular vibrations and thermal disorder are ultimately reflected in the mobility of the materials. Based on the transfer integrals alone, alkylation of DNTT should therefore boost the mobility by a factor of 2.5: in DNTT the combination of $J_x$ has the wrong sign, while it becomes ideally isotropic in alkylated DNTT[15]. However, the enhancement of thermal disorder by the long-axis sliding 'killer' mode is so strong that it cancels much of this expected gain; as a result alkylated DNTT is predicted to have only about 30% higher mobility than DNTT, which is consistent with experimental observations on single crystals. An analogous compensation is predicted for the other pair of molecules, BTBT and C8-BTBT-C8. Following the same line of reasoning, the exceptionally high mobility value of rubrene can be attributed to its small energetic disorder due to its reduced sensitivity to the long-axis sliding motion, which compensates for its unfavourable anisotropic transfer integrals. In this material an extra boost comes from two secondary factors: its large lattice parameters and, somewhat counterintuitively, a comparatively small value of the bandwidth scale $J$, which ensures a sufficient thermal population of higher energy and more delocalized states in its quasi 1D-bands[50]. In pentacene the unfavorable combination of $J_x$, the higher disorder compared to rubrene or DNTT as well as unfavourable secondary factors all concur to make it the worst performer in the set.

What is striking from these considerations is that none of the state-of-the-art molecules considered here is ideal in the two aspects that have the largest impact on the charge mobility, i.e. none of them combines low energetic disorder with isotropic transfer integrals. If it were possible to rationally design molecules that meet these requirements simultaneously this could enable molecular semiconductors with very high mobilities, potentially exceeding that of rubrene. In this respect, we have shown here that the alkylated compounds, which were selected as very promising candidates owing to their isotropic electronic properties, are in fact hampered by the presence of very strong disorder and that such disorder is carried almost entirely by a single inter-molecular sliding mode. The beneficial reduction of the amplitude of the sliding fluctuations achieved from a stiffer structure is entirely compensated by a concomitant large increase of their coupling to electrons, enhancing their detrimental impact on transport. The origin of this degradation immediately becomes transparent when crossing the information on the sliding mode dependence of transfer integrals (see in Figure 3k) with that on supramolecular packing (see Table S7). In the unsubstituted DNTT and BTBT crystals, the tilt angle of the molecules with respect to the herringbone plane induces an ideal offset in the relative alignment of adjacent molecules at equilibrium: the transfer integrals are near their extremum, where the coupling of electrons to the sliding mode vanishes. The proximity to this local "sweet spot" is lost upon alkylation, as the tilting angle is reduced.

In terms of rational molecular design guidelines for new high mobility materials the general message emerging from the present analysis is that intermolecular shifts at equilibrium determine not only the set of transfer integrals $J_x$ but also their dependence on the long-axis sliding, i.e. they ultimately control *both* ingredients that govern charge transport in 2D herringbone structures. Since such relative shifts are directly affected by the molecular tilting angle, the latter emerges as a natural parameter to play with in the quest for higher mobility semiconductors. This could be achieved by the engineering of side chains, variation of their length, branching or incorporation of specific binding sites. A second important lesson that follows from our insight is that a proper screening of promising candidates in molecular crystal structure databases can be achieved with relatively simple quantum chemistry methods, without the need of performing expensive solid-state phonon calculations. Indeed, evaluating the set of $J_x$ and the corresponding $dJ_x/dz$ already contains the key figures needed to rationalize transport.

# Methods

## 1. Materials

BTBT and C8-BTBT-C8 were synthesized according to previously described procedures[51–54]. Rubrene and DNTT were purchased from Sigma-Aldrich and used as received. Pentacene was purchased from TCI and used as received. C8-DNTT-C8 was supplied by Nippon Kayaku.

## 2. Terahertz Time-Domain Spectroscopy (THz-TDS)

All samples were prepared for THz-TDS measurements by first mixing the investigated material (around 50 mg of polycrystalline powder) with polyethylene (average molecular weight of 35000 amu, Sigma-Aldrich, Poole, UK) to a concentration of 10% w/w and grinding with a mortar and pestle to homogenize the mixture and reduce particle size. The samples were then pressed into 13 x 2 mm (diameter x height) freestanding pellets using a hydraulic press under 2 tons. A corresponding pellet containing only polyethylene was made with the same dimensions to act as a standard blank. All THz-TDS measurements were performed using a commercial Terapulse 4000 (Teraview Ltd, Cambridge, UK), fitted with a Janis liquid nitrogen cryostat (Wobourn, Massachusetts, USA). Each measurement is comprised of 4000 averaged time-domain waveforms, which was subsequently Fourier transformed to yield a terahertz power spectrum. The terahertz absorption spectra presented were generated through division of a sample and blank spectra.

## 3. Specular X-ray diffraction (sXRD) and crystal structure solving for C8-DNTT-C8

A polycrystalline sample of C8-DNTT-C8 was loaded into a 0.7 mm borosilicate glass capillary and transmission PXRD data were collected over the range 3–80° 2θ (2 kW; Cu Kα$_1$, 1.54056 A°; step size 0.017° 2θ), for a total of 22 hrs using a variable count time scheme. The Bruker D8 Advance diffractometer was equipped with a LynxEye detector. The data indexed to a monoclinic cell (a = 33.543 Å, b = 7.866 Å, c = 5.983 Å, β = 90.35°, V =

1578.6 Å$^3$), were Pawley fitted in *DASH*[55] and then the most probable space group of *P*2$_1$/*c* determined by use of *ExtSym*.[56] Consideration of the cell volume, molecular volume and space group symmetry suggested Z′ = 0.5. The simulated annealing component of *DASH* was used to optimise the orientation and conformation of a half-molecule model of C8-DNTT-C8 against the diffraction data (107 reflections), with the molecular centre of symmetry anchored on the origin of the unit cell. A refineable (1 0 0) correction for preferred orientation was included at the structure solution stage. This yielded a favourable $\chi^2_{SA}/\chi^2_{Pawley}$ ratio of around 3 for the best solution, with a chemically and crystallographically sensible structure. The validity of the structure was confirmed by a Rietveld refinement against data in the range 4–60° 2θ using *TOPAS*[57] and by fixed-cell energy minimisation of the crystal structure using DFT-D. The energy-minimised structure was then used as the basis of a more accurate starting model of a half molecule, and the structure then re-solved and re-refined in space group *P*2$_1$/*a* to facilitate comparison with other related structures. The final Rietveld refinement, using a spherical harmonics correction for preferred orientation, yield an $R_{wp}$ value of 3.6 (c.f. Pawley $R_{wp}$ of 1.8).

## 4. Inelastic Neutron Scattering (INS)

High-resolution INS measurements on a powder sample of 2g of BTBT, 1g of C8-BTBT-C8, 1g of DNTT, 1g of C8-DNTT-C8, 1g of rubrene and 1g of pentacene respectively, contained in a flat aluminum cell of cross-sectional area 4×4.8 cm$^2$ were performed at the inverted-geometry neutron spectrometer TOSCA, spanning energy transfers up to 4000 cm$^{-1}$, at the ISIS Pulsed Neutron & Muon Source, Rutherford Appleton Laboratory, United Kingdom[58–60]. The spectral resolution of TOSCA is similar to that of infrared and Raman techniques, approximately 2 % of the energy in question. The sample was studied at a temperature of ~10 K using a dedicated closed-cycle helium refrigerator, so as to minimize the effect of the Debye Waller factor, thus allowing us to fully distinguish the individual peaks.[61]

## 5. Organic Field-Effect Transistors (OFETs)

**Materials**: Octadecyltrimethoxysilane (OTMS) was purchased from Fluorochem and used as received (storage under a $N_2$ atmosphere to prevent hydrolysis). Highly doped n-type Si (100) wafers (resistivity < 0.005 Ω cm, Active Business Company GmbH) were used as the substrates for thin film transistors fabrication. A 300 nm $SiO_2$ layer (capacitance $C_i$ = 11.5 nF cm$^{-2}$) was thermally grown onto the Si substrates as a gate dielectric.

**Field-Effect Transistor Fabrication**: 30 nm films of DNTT and C8-DNTT-C8 were deposited by thermal evaporation (Angstrom Engineering, Inc.) at a rate of 0.2 Å s$^{-1}$ and 0.1 Å s$^{-1}$ respectively, under pressure of 10$^{-7}$ Torr and substrate rotation of 25 RPM, onto octadecyltrimethoxysilane (OTMS) treated 300 nm thermally grown $SiO_2$ substrates. The monolayer was deposited according to a method published previously[62]. The substrate temperature was maintained at room temperature during the deposition process. Approximately 40 nm of gold were thermally evaporated through a shadow mask (L = 350 μm, W = 1000 μm) at a rate of 0.5 Å s$^{-1}$ to complete the devices.

**Device Characterization**: Transistors were characterized using an Agilent 4155B Semiconductor Parameter Analyser and standard probe station setup at room temperature in a $N_2$-filled glovebox. Device parameters were extracted using the standard calculation techniques. The devices were stored in the dark in a $N_2$-filled glovebox.

**Reliability Factor (r)**[44]: is designed to show an overall ability of a reported OFET to "deliver on its promise". This parameter is defined as the ratio of the maximum channel conductivity experimentally achieved in a FET to the maximum channel conductivity expected in the correctly functioning transistor with the claimed carrier mobility,
$$r \equiv \frac{\sigma_{achieved}^{max}}{\sigma_{well-behaved}^{max}}$$. The parameter $r$ simply shows if the reported μ is a good metrics of the actual electrical performance of a FET.

**6. Density Functional Theory Simulations of vibrational modes**

The periodic solid-state density functional theory (DFT) simulations were performed using a developmental version of the CRYSTAL17 software package.[63] The Grimme D3-corrected[36,37] Perdew-Burke-Ernzerhof (PBE) density functional[38] was coupled with the split-valence triple-zeta 6-311G(2d,2p) basis set[64] for all atoms. All the simulations were initiated from the experimental single-crystal X-ray diffraction (SCXRD) determined structures,[52,65–68] with the exception of C8-DNTT-C8 which was determined for the first time for this study using a combination of DFT simulations and refinement against the experimental powder X-ray diffraction (PXRD) data. All structures were fully optimized (atomic positions and lattice vectors) with no constraints other than the space group symmetry of the solid. Subsequently, frequency calculations were performed via numerical displacements within the harmonic approximation, and IR-intensities were determined via the Berry phase method.[69–71]

In most cases, the harmonic approximation was sufficient for reproducing the vibrational potential energy curves for the OSC normal modes, however in the case of the alkylated materials there existed significant anharmonicity (deviation from harmonic behavior) in the lowest frequency modes (<30 cm$^{-1}$). In these modes, the explicit anharmonic vibrational energies were determined by scanning the individual normal mode coordinates and determining the explicit vibrational potential energy curves. These values were then fitted with a sixth-order polynomial, and the one-dimensional anharmonic oscillator Schrodinger equation was used to solve for the vibrational energies. The energy convergence criteria were set to $\Delta E < 10^{-8}$ and $10^{-10}$ hartree for the optimization and vibrational calculations, respectively.

## 7. Calculation of electron-phonon coupling constants and charge carrier mobilities

Local (Holstein) and nonlocal (Peierls) linear electron-phonon coupling calculations[7] in periodic solids were performed following the protocol by Girlando et al.[42] with two important improvements: (i) we rely on a full *ab initio* description of vibrational modes and their coupling to electrons; (ii) we explicitly account for the anharmonicity that strongly affects low-frequency modes of alkylated systems. DFT calculations (PBE0 functional, def2-SVP basis, ORCA v4.0 code)[72] were performed as a function of the displacement along each mode to compute site energies (HOMO level) and transfer integrals for the symmetry-inequivalent molecular pairs. Transfer integrals are computed as one-electron HOMO-HOMO couplings following the dimer projection approach.[73]

Charge transport simulations within the transient localization framework in the relaxation time approximation[14] were performed as described in a previous paper,[15] for a two-dimensional tight-binding model for hole states fully parametrized with *ab initio* inputs for the specific systems at hand. We have considered lattices of 320,000 molecules as to ensure the convergence of the calculated mobility values at 300K. The numerical propagation of carriers' wavefunctions is efficiently performed with the technique introduced in a previous reference.[74]

## Supporting Information

Supporting Information is available from the Wiley Online Library or from the author.

## Acknowledgements

The authors gratefully acknowledge the ISIS neutron and muon source for beam time (TOSCA instrument). G.S. acknowledges postdoctoral fellowship support from the Wiener-Anspach Foundation and The Leverhulme Trust (Early Career Fellowship supported by the Isaac Newton Trust). D.J.H. acknowledges the EPSRC Centre for Doctoral Training in Plastic Electronics EP/G037515/1. G.L. thanks the support from the Royal Society (Newton Fellowship, NF151515). Financial support from the German Research Foundation (BR4869-1/1) is gratefully acknowledged (K.B.). We gratefully acknowledge support from the European Research Council (ERC, Synergy Grant 610115) and the Engineering and Physical Sciences Research Council (EPSRC, programme grant EP/M005143/1). M.T.R. and J.A.Z. thank the Engineering and Physical Sciences Research Council (EPSRC, programme grant EP/N022769/1), and computational resources via membership of the UK's HEC Materials Chemistry Consortium (funded by the EPSRC, programme grant EP/L000202) and the ARCHER UK National Supercomputing Service (http://www.archer.ac.uk). The authors thank Nippon Kayaku for supplying C8-DNTT-C8. Y.H.G. thanks the Belgian National Fund for Scientific Research (FNRS - Research Fellow PhD grant for A.R. and project BTBT no. 2.4565.11) and the Walloon Region (WCS project 1117306). The University of Reading's Chemical Analysis Facility is acknowledged for access to the Bruker D8 Advance diffractometer. G.D. and S.F. thank S. Ciuchi, A. Girlando and A. Brillante for useful discussions. G.D. acknowledges computational resources provided by GENCI-CINES/IDRIS.

**Author contributions**

G.S. and M.T.R. conceived and designed the experiments. D.J.H. and K.B. contributed to the realization of the project. M.T.R. and J.A.Z. performed the THz-TDS experiments and data analysis. G.S., M.T.R., G.L. and J.A. carried out INS measurements and data analysis. D.V. designed the devices architecture. G.S. fabricated the devices and performed the TLM measurements. A.R., C.R. and Y.H.G. synthesized and supplied the BTBT derivatives. K.T. provided the DNTT derivatives. A.R.K. and K.S. solved the crystal structure of C8-DNTT-C8. G.D. and S.F. realized the electron-phonon coupling and the transient localization mobility calculations. G.S., G.D. and M.T.R. wrote the first draft of the manuscript. All authors discussed the results and reviewed the manuscript. S.F. and H.S. supervised the project.

## Conflict of Interest

The authors declare no conflict of interest.

## Additional information

Correspondence and requests for materials should be addressed to G.S., S.F. or H.S.

## Keywords

Organic electronics, charge transport, dynamic disorder, molecular design

# References

[1] H. Sirringhaus, *Adv. Mater.* **2014**, *26*, 1319.

[2] M. Nikolka, I. Nasrallah, B. Rose, M. K. Ravva, K. Broch, A. Sadhanala, D. Harkin, J. Charmet, M. Hurhangee, A. Brown, S. Illig, P. Too, J. Jongman, I. McCulloch, J. L. Bredas, H. Sirringhaus, *Nat. Mater.* **2017**, *16*, 356.

[3] G. Schweicher, Y. Olivier, V. Lemaur, Y. H. Geerts, *Isr. J. Chem.* **2014**, *54*, 595.

[4] Y. Yamashita, F. Hinkel, T. Marszalek, W. Zajaczkowski, W. Pisula, M. Baumgarten, H. Matsui, K. Müllen, J. Takeya, *Chem. Mater.* **2016**, *28*, 420.

[5] C. Mitsui, T. Okamoto, M. Yamagishi, J. Tsurumi, K. Yoshimoto, K. Nakahara, J. Soeda, Y. Hirose, H. Sato, A. Yamano, T. Uemura, J. Takeya, *Adv. Mater.* **2014**, *26*, 4546.

[6] V. Podzorov, E. Menard, A. Borissov, V. Kiryukhin, J. A. Rogers, M. E. Gershenson, *Phys. Rev. Lett.* **2004**, *93*, 1.

[7] V. Coropceanu, J. Cornil, D. A. da Silva Filho, Y. Olivier, R. Silbey, J. L. Brédas, *Chem. Rev.* **2007**, *107*, 926.

[8] N. Boden, R. J. Bushby, J. Clements, B. Movaghar, K. J. Donovan, T. Kreouzis, *Phys. Rev. B* **1995**, *52*, 13274.

[9] N. E. Lee, J. J. Zhou, L. A. Agapito, M. Bernardi, *Phys. Rev. B* **2018**, *97*, 1.

[10] A. S. Eggeman, S. Illig, A. Troisi, H. Sirringhaus, P. a Midgley, *Nat. Mater.* **2013**, *12*, 1045.

[11] S. Illig, A. S. Eggeman, A. Troisi, L. Jiang, C. Warwick, M. Nikolka, G. Schweicher, S. G. Yeates, Y. Henri Geerts, J. E. Anthony, H. Sirringhaus, *Nat. Commun.* **2016**, *7*, 1.

[12] A. Troisi, G. Orlandi, *J. Phys. Chem. A* **2006**, *110*, 4065.

[13] A. Troisi, G. Orlandi, *Phys. Rev. Lett.* **2006**, *96*, 1.

[14] S. Fratini, D. Mayou, S. Ciuchi, *Adv. Funct. Mater.* **2016**, *26*, 2292.

[15] S. Fratini, S. Ciuchi, D. Mayou, G. T. De Laissardière, A. Troisi, *Nat. Mater.* **2017**, *16*, 998.

[16] J. F. Chang, T. Sakanoue, Y. Olivier, T. Uemura, M. B. Dufourg-Madec, S. G. Yeates, J. Cornil, J. Takeya, A. Troisi, H. Sirringhaus, *Phys. Rev. Lett.* **2011**, *107*, 1.

[17] T. Uemura, M. Yamagishi, J. Soeda, Y. Takatsuki, Y. Okada, Y. Nakazawa, J. Takeya, *Phys. Rev. B - Condens. Matter Mater. Phys.* **2012**, *85*, 1.

[18] K. Sakai, Y. Okada, T. Uemura, J. Tsurumi, R. Häusermann, H. Matsui, T. Fukami, H. Ishii, N. Kobayashi, K. Hirose, J. Takeya, *NPG Asia Mater.* **2016**, *8*, e252.

[19] T. Kubo, R. Häusermann, J. Tsurumi, J. Soeda, Y. Okada, Y. Yamashita, N. Akamatsu, A. Shishido, C. Mitsui, T. Okamoto, S. Yanagisawa, H. Matsui, J. Takeya, *Nat. Commun.* **2016**, *7*, 11156.

[20] S. Ciuchi, R. C. Hatch, H. Höchst, C. Faber, X. Blase, S. Fratini, *Phys. Rev. Lett.* **2012**, *108*, 1.

[21] F. Bussolotti, J. Yang, T. Yamaguchi, K. Yonezawa, K. Sato, M. Matsunami, K. Tanaka, Y. Nakayama, H. Ishii, N. Ueno, S. Kera, *Nat. Commun.* **2017**, *8*, DOI 10.1038/s41467-017-00241-z.

[22] W. Xie, K. a. McGarry, F. Liu, Y. Wu, P. P. Ruden, C. J. Douglas, C. D. Frisbie, *J. Phys. Chem. C* **2013**, *117*, 11522.

[23] Y. S. Aytekin, M. Köktürk, A. Zaczek, T. M. Korter, E. J. Heilweil, O. Esenturk, *Chem. Phys.* **2018**, *512*, 36.

[24] M. T. Ruggiero, J. A. Zeitler, A. Erba, *Chem. Commun.* **2017**, *53*, 3781.

[25] M. T. Ruggiero, J. J. Sutton, S. J. Fraser-Miller, A. J. Zaczek, T. M. Korter, K. C. Gordon, J. A. Zeitler, *Cryst. Growth Des.* **2018**, *18*, 6513.

[26] M. T. Ruggiero, W. Zhang, A. D. Bond, D. M. Mittleman, J. A. Zeitler, *Phys. Rev.*

*Lett.* **2018**, *120*, 196002.

[27] A. J. Zaczek, L. Catalano, P. Naumov, T. M. Korter, *Chem. Sci.* **2019**, DOI 10.1039/C8SC03897J.

[28] A. Brillante, I. Bilotti, R. G. Della Valle, E. Venuti, M. Masino, A. Girlando, *Adv. Mater.* **2005**, *17*, 2549.

[29] A. Brillante, I. Bilotti, R. G. Della Valle, E. Venuti, A. Girlando, *CrystEngComm* **2008**, *10*, 937.

[30] N. Bedoya-Martínez, B. Schrode, A. O. F. Jones, T. Salzillo, C. Ruzié, N. Demitri, Y. H. Geerts, E. Venuti, R. G. Della Valle, E. Zojer, R. Resel, *J. Phys. Chem. Lett.* **2017**, *8*, 3690.

[31] Z. Tu, Y. Yi, V. Coropceanu, J. L. Brédas, *J. Phys. Chem. C* **2018**, *122*, 44.

[32] T. F. Harrelson, V. Dantanarayana, X. Xie, C. Koshnick, D. Nai, R. Fair, S. A. Nuñez, A. K. Thomas, T. L. Murrey, M. A. Hickner, J. K. Grey, J. E. Anthony, E. D. Gomez, A. Troisi, R. Faller, A. J. Moulé, *Mater. Horizons* **2018**, DOI 10.1039/C8MH01069B.

[33] W. Xie, K. Willa, Y. Wu, R. Häusermann, K. Takimiya, B. Batlogg, C. D. Frisbie, *Adv. Mater.* **2013**, *25*, 3478.

[34] K. Nakayama, Y. Hirose, J. Soeda, M. Yoshizumi, T. Uemura, M. Uno, W. Li, M. J. Kang, M. Yamagishi, Y. Okada, E. Miyazaki, Y. Nakazawa, A. Nakao, K. Takimiya, J. Takeya, *Adv. Mater.* **2011**, *23*, 1626.

[35] N. Bedoya-Martínez, A. Giunchi, T. Salzillo, E. Venuti, R. G. Della Valle, E. Zojer, *J. Chem. Theory Comput.* **2018**, *14*, 4380.

[36] S. Grimme, J. Antony, S. Ehrlich, H. Krieg, *J. Chem. Phys.* **2010**, *132*, DOI 10.1063/1.3382344.

[37] S. Grimme, S. Ehrlich, L. Goerigk, *J. Comput. Chem.* **2011**, *32*, 1456.

[38] J. P. Perdew, K. Burke, M. Ernzerhof, *Phys. Rev. Lett.* **1996**, *77*, 3865.

[39]  I. Natkaniec, E. L. Bokhenkov, B. Dorner, J. Kalus, G. A. MacKenzie, G. S. Pawley, U. Schmelzer, E. F. Sheka, *J. Phys. C Solid State Phys.* **1980**, *13*, 4265.

[40]  N. Vukmirović, C. Bruder, V. M. Stojanović, *Phys. Rev. Lett.* **2012**, *109*, 1.

[41]  R. S. Sánchez-Carrera, P. Paramonov, G. M. Day, V. Coropceanu, J. L. Brédas, *J. Am. Chem. Soc.* **2010**, *132*, 14437.

[42]  A. Girlando, L. Grisanti, M. Masino, I. Bilotti, A. Brillante, R. G. Della Valle, E. Venuti, *Phys. Rev. B - Condens. Matter Mater. Phys.* **2010**, *82*, 1.

[43]  D. a. Da Silva Filho, E. G. Kim, J. L. Brédas, *Adv. Mater.* **2005**, *17*, 1072.

[44]  H. H. Choi, K. Cho, C. D. Frisbie, H. Sirringhaus, V. Podzorov, *Nat. Mater.* **2018**, *17*, 2.

[45]  K. Pei, M. Chen, Z. Zhou, H. Li, P. K. L. Chan, *ACS Appl. Electron. Mater.* **2019**, *1*, 379.

[46]  M. Dohr, O. Werzer, Q. Shen, I. Salzmann, C. Teichert, C. Ruzié, G. Schweicher, Y. H. Geerts, M. Sferrazza, R. Resel, *Chemphyschem* **2013**, *14*, 2554.

[47]  M. Dohr, H. M. A. Ehmann, A. O. F. Jones, I. Salzmann, Q. Shen, C. Teichert, C. Ruzié, G. Schweicher, Y. H. Geerts, R. Resel, M. Sferrazza, O. Werzer, *Soft Matter* **2017**, *13*, 2322.

[48]  R. Janneck, N. Pilet, S. P. Bommanaboyena, B. Watts, P. Heremans, J. Genoe, C. Rolin, *Adv. Mater.* **2017**, *29*, 1.

[49]  Y. Yi, V. Coropceanu, J. L. Brédas, *J. Chem. Phys.* **2012**, *137*, DOI 10.1063/1.4759040.

[50]  S. Ciuchi, S. Fratini, *Phys. Rev. B - Condens. Matter Mater. Phys.* **2012**, *86*, 1.

[51]  M. Saito, I. Osaka, E. Miyazaki, K. Takimiya, H. Kuwabara, M. Ikeda, *Tetrahedron Lett.* **2011**, *52*, 285.

[52]  H. Ebata, T. Izawa, E. Miyazaki, K. Takimiya, M. Ikeda, H. Kuwabara, T. Yui, *J. Am.*

*Chem. Soc.* **2007**, *129*, 15732.

[53] M. J. Kang, T. Yamamoto, S. Shinamura, E. Miyazaki, K. Takimiya, *Chem. Sci.* **2010**, *1*, 179.

[54] M. J. Kang, I. Doi, H. Mori, E. Miyazaki, K. Takimiya, M. Ikeda, H. Kuwabara, *Adv. Mater.* **2011**, *23*, 1222.

[55] W. I. F. David, K. Shankland, J. Van De Streek, E. Pidcock, W. D. S. Motherwell, J. C. Cole, *J. Appl. Crystallogr.* **2006**, *39*, 910.

[56] A. J. Markvardsen, K. Shankland, W. I. F. David, J. C. Johnston, R. M. Ibberson, M. Tucker, H. Nowell, T. Griffin, *J. Appl. Crystallogr.* **2008**, *41*, 1177.

[57] A. A. Coelho, *J. Appl. Crystallogr.* **2018**, *51*, 210.

[58] S. F. Parker, F. Fernandez-Alonso, A. J. Ramirez-Cuesta, J. Tomkinson, S. Rudić, R. S. Pinna, G. Gorini, J. F. Castanon, *J. Phys. Conf. Ser.* **2014**, *554*, 012003.

[59] R. S. Pinna, S. Rudić, S. F. Parker, G. Gorini, F. Fernandez-Alonso, *EPJ Web Conf.* **2015**, *83*, 1.

[60] R. S. Pinna, S. Rudić, S. F. Parker, J. Armstrong, M. Zanetti, G. Škoro, S. P. Waller, D. Zacek, C. A. Smith, M. J. Capstick, D. J. McPhail, D. E. Pooley, G. D. Howells, G. Gorini, F. Fernandez-Alonso, *Nucl. Instruments Methods Phys. Res. Sect. A Accel. Spectrometers, Detect. Assoc. Equip.* **2018**, *896*, 68.

[61] K. Dymkowski, S. F. Parker, F. Fernandez-Alonso, S. Mukhopadhyay, *Phys. B Condens. Matter* **2018**, DOI 10.1016/j.physb.2018.02.034.

[62] Y. Ito, A. A. Virkar, S. Mannsfeld, H. O. Joon, M. Toney, J. Locklin, Z. Bao, *J. Am. Chem. Soc.* **2009**, *131*, 9396.

[63] R. Dovesi, A. Erba, R. Orlando, C. M. Zicovich-Wilson, B. Civalleri, L. Maschio, M. Rérat, S. Casassa, J. Baima, S. Salustro, B. Kirtman, *Wiley Interdiscip. Rev. Comput. Mol. Sci.* **2018**, e1360.

[64] R. Krishnan, J. S. Binkley, R. Seeger, J. A. Pople, *J. Chem. Phys.* **1980**, *72*, 650.

[65] O. D. Jurchescu, A. Meetsma, T. T. M. Palstra, *Electrochem. Solid-State Lett.* **2006**, *9*, 330.

[66] D. Holmes, S. Kumaraswamy, A. J. Matzger, K. P. C. Vollhardt, *Chem. Eur. J.* **1999**, *5*, 3399.

[67] T. Yamamoto, K. Takimiya, *J. Am. Chem. Soc.* **2007**, *129*, 2224.

[68] C. Niebel, Y. Kim, C. Ruzié, J. Karpinska, B. Chattopadhyay, G. Schweicher, A. Richard, V. Lemaur, Y. Olivier, J. Cornil, A. R. Kennedy, Y. Diao, W.-Y. Lee, S. Mannsfeld, Z. Bao, Y. H. Geerts, *J. Mater. Chem. C* **2015**, *3*, 674.

[69] C. M. Zicovich-Wilson, F. Pascale, C. Roetti, V. R. Saunders, R. Orlando, R. Dovesi, *J. Comput. Chem.* **2004**, *25*, 1873.

[70] F. Pascale, C. M. Zicovich-Wilson, F. López Gejo, B. Civalleri, R. Orlando, R. Dovesi, *J. Comput. Chem.* **2004**, *25*, 888.

[71] Y. Noel, C. Zicovich-Wilson, B. Civalleri, P. D'Arco, R. Dovesi, *Phys. Rev. B* **2001**, *65*, 1.

[72] F. Neese, *Wiley Interdiscip. Rev. Comput. Mol. Sci.* **2012**, *2*, 73.

[73] E. F. Valeev, V. Coropceanu, D. A. Da Silva Filho, S. Salman, J. L. Brédas, *J. Am. Chem. Soc.* **2006**, *128*, 9882.

[74] F. Triozon, J. Vidal, R. Mosseri, D. Mayou, *Phys. Rev. B - Condens. Matter Mater. Phys.* **2002**, *65*, 2202021.